# Full Eye Length Spectral Domain OCT with Single-Modulator Depth Multiplexing

**XUE DONG[1], ASTER ZHANG[2,3], FABIO FEROLDI[1,4], SARAH SINGH[1], AUSTIN ROORDA[1], GUANGHAN MENG[1,2,†,*], AND LAURA WALLER[2,†]**

*[1]Herbert Wertheim School of Optometry and Vision Science, University of California, Berkeley, Berkeley, CA, 94720*
*[2]Department of Electrical Engineering and Computer Science, University of California, Berkeley, Berkeley, CA, 94720*
*[3]Current address: Institute for Research in Electronics and Applied Physics, University of Maryland, College Park, MD, 20742*
*[4]Current address: Taara Connect, Inc., Sunnyvale, CA 94089*
*†These authors contributed equally to this research*
**Correspondence: guanghan_meng@berkeley.edu (G.M.)*

**Abstract:** Optical coherence tomography (OCT) systems generally face a tradeoff between imaging depth range and system cost: swept-source OCT (SS-OCT) enables long imaging ranges at high cost, whereas spectral-domain OCT (SD-OCT) is more affordable but cannot capture the full axial length of the eye in a single acquisition. Depth-multiplexing (DM) techniques extend SD-OCT depth range by capturing images at multiple depths simultaneously, but existing approaches require multiple light modulation units or detectors, partially negating the cost advantage of SD-OCT. Here we present a DM-SD-OCT system that achieves depth multiplexing with a single light modulation unit and a single line detector. Signals from three distinct depths are captured simultaneously, with the single modulation unit encoding each signal with a distinct phase ramp for computational demixing. Our method is demonstrated on *in vivo* human eyes, providing a practical, novel and relatively low-cost solution for extended-range OCT imaging.

## 1. INTRODUCTION

Optical coherence tomography (OCT) is a high-resolution optical imaging technique that generates depth-resolved cross-sectional images using interferometry [1]. As the optical counterpart of ultrasound, OCT detects echoes of light and provides micrometer-scale depth resolution in living tissue [1]. In the eye clinic, OCT is routinely used for the diagnosis and monitoring of retinal diseases [2]. OCT also serves as a biometer, quantifying ocular dimensions such as the axial length, i.e., the distance between the anterior corneal surface and the photoreceptor layer of the retina [3]. Axial length is the primary metric for tracking eye elongation during myopia progression, a growing concern given that myopia is projected to affect half the world's population by 2050 [4], and for identifying abnormal eye growth in pediatric patients before visual consequences emerge [5].

Measuring the full axial length of the eye requires an extended imaging range. Adult axial length averages 23-24 mm, while high myopia is associated with axial lengths exceeding 26 mm [6]. In Fourier domain OCT, i.e., swept-source OCT (SS-OCT) and spectral-domain OCT (SD-OCT), the depth profile is obtained via an inverse Fourier transform of the measured spectrum. The axial imaging range is therefore inversely proportional to the spectral sampling interval. SS-OCT is well suited for axial length measurement because it samples the optical spectrum sequentially with fine spectral increments using a wavelength-swept laser, such as a MEMS-tuned external cavity tunable laser (ECTL) or vertical cavity surface emitting laser (VCSEL) [7]. This dense spectral sampling provides the extended imaging range required for full eye length imaging. As a result, SS-OCT is commonly used for ocular biometry and is considered the gold standard. However, SS-OCT systems are expensive, often more than twice the cost of an SD-OCT in the clinic, largely because of the high-cost light source and the high-bandwidth data acquisition system. Recently, thermally tuned VCSELs have reduced source cost for full-eye SS-OCT [8], yet the high-bandwidth acquisition requirement remains.

SD-OCT, in contrast, records the spectrum simultaneously using a broadband light source and spectrometer, making it more cost-effective and widely accessible. However, its imaging range is limited by the spectrometer resolution to a few millimeters, which makes full-eye-length imaging challenging. One strategy to circumvent this limitation is to integrate multiple reference arms targeting different depths within the eye. Previous approaches fall into two categories: sequential acquisition and simultaneous acquisition. Sequential acquisition alternates between reference arms [9–12], which sacrifices temporal resolution and leaves the measurement vulnerable to eye motion. Simultaneous acquisition methods integrate multiple interferometers or optical modulation devices, one per reference arm [13,14], increasing cost and system complexity and thus compromising accessibility. A single-interferometer depth-multiplexing strategy has been developed for ECTL-based SS-OCT [15], but this method relies on coherence revival, a periodic reappearance of interference fringes at path length differences matching multiples of the laser's cavity length. This method therefore does not transfer to SD-OCT, whose broadband sources lack a laser cavity.

Here, we present a computational depth-multiplexing approach that extends the imaging depth of SD-OCT using a single interferometer and simultaneous acquisition. The technique employs a single scanning mirror to encode distinct phase ramps for multiple reference arms, enabling full eye length imaging in a single scan. As all the light splitting and modulation occurs in the reference arm, the system collects signal returning from the sample arm without any additional loss, which is critical as ocular safety standards cap the optical power that can be delivered to the eye. We demonstrate that our photon-efficient, single-modulator depth multiplexing approach enables simultaneous acquisition and computational demixing of signals from three depths within the eye: cornea, pupil, and retina, and validate axial length measurements in both emmetropic and myopic subjects.

## 2. METHODS

### A. Principle and Theory of the Single-Modulator Depth-Multiplexing (DM) SD-OCT

Conventional SD-OCT systems operate by splitting a broadband light source into a sample arm and a single, stationary reference arm. The light backscattered from the eye interferes with the reference reflection, and the resulting spectral interferogram is captured by a spectrometer comprising a diffraction grating and a high-speed line camera. Applying an inverse Fourier transform to the measured interferogram yields a depth-dependent axial reflectivity profile (A-scan). By laterally scanning the sample arm beam across the eye, adjacent A-scans are acquired to form a two-dimensional cross-sectional image, termed a B-scan. However, because the detected interferogram is real-valued and lacks the complex phase information required to distinguish positive from negative optical path delay, the inverse Fourier transform produces a mirror image (a complex conjugate artifact) about the zero-delay position for each A-scan. In addition, the maximum achievable axial imaging range is fundamentally limited by the finite number of pixels in the line camera.

Our depth-multiplexing (DM) SD-OCT method draws on a phase-modulation principle previously used to remove the complex conjugate artifact in OCT image reconstruction [16–18]. By deviating the light beam slightly off-center from the pivot axis of the galvanometer (galvo) scanner (**Fig. 1A**), an optical path length (OPL) change is introduced as the galvo scans around its rotation axis, resulting in a linear phase ramp along the lateral-scanning direction (transverse axis) of the B-scan. Applying a Fourier transform (FT) to the B-scan signal along the transverse direction separates the signal and its complex conjugate into opposite quadrants of the Fourier domain. Cropping the signal from one quadrant yields a complex-valued representation, from which both amplitude and phase components can be recovered. This complex representation suppresses the complex conjugate artifact during image reconstruction. The offset between the beam and the pivot axis of the galvo determines the slope of the phase ramp and therefore the location of the signal in the Fourier domain. It follows that if multiple beams are simultaneously introduced onto the galvo at different offsets from the center, their corresponding Fourier domain signals will be shifted to distinct locations, allowing them to be separated computationally in reconstruction.

For our full-length eye imaging approach, three reference arms targeting the cornea, pupil, and retina are simultaneously modulated by a second galvo introduced to the reference arm set (referred to here as the *reference scanner*) and imaged with the same interferometer (**Fig. 2**). The reference scanner is synchronized with the sample scanner, and the three reference beams with varying OPLs strike it with different offsets from the pivot axis (**Fig. 1B**), resulting in distinct phase ramps along the transverse axis of the B-scan. A single scanner now encodes all three reference arms, and the three depths can be demixed computationally.

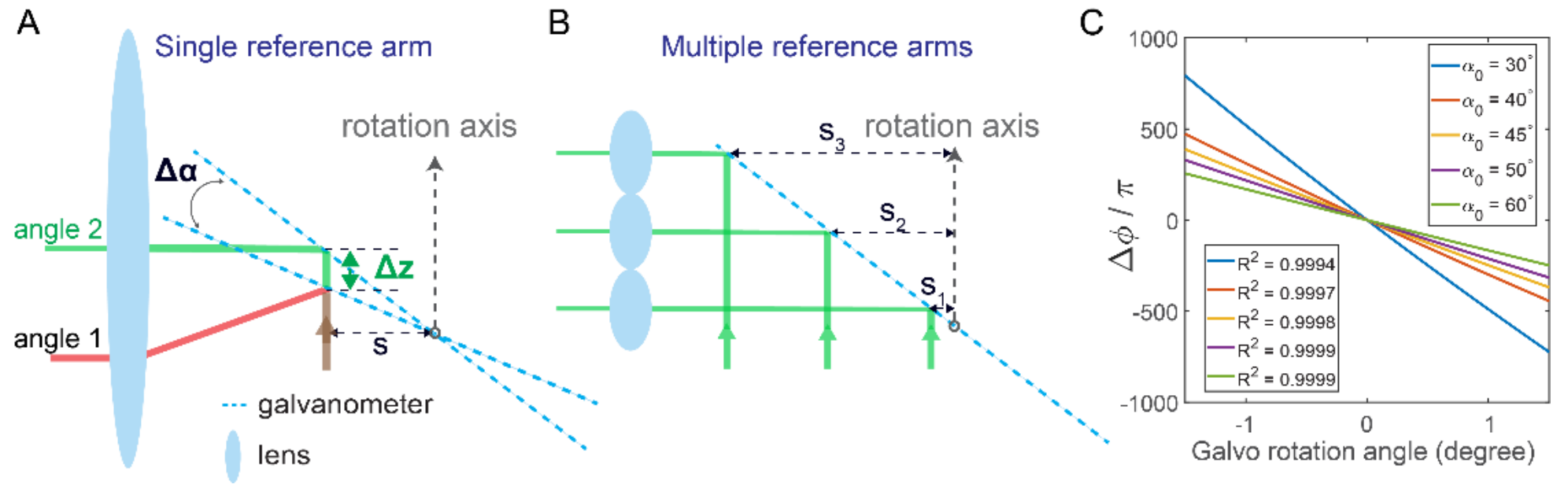

**Figure 1. Principle of scan-induced phase modulation. (A)** Beam geometry for a single reference arm with phase modulation. The incident beam is laterally displaced by a distance $s$ from the galvo mirror's rotation axis. As the mirror rotates through a mechanical angle Δα, the beam–mirror intersection point shifts, changing the optical path length (OPL) by Δz. **(B)** Beam geometry for multiple reference arms with a single phase modulator. Different offsets $s_1$ $s_2$ and $s_3$ produce distinct slopes of phase shifts in the 3 reference arms. **(C)** Simulation showing how scan-induced phase modulation varies with the central angle of incidence (AOI) at the galvo, $\alpha_0$. The phase shift is plotted against the galvo rotation angle relative to its central position, Δα, for $\alpha_0$ values of 30°, 40°, 45°, 50°, and 60°. The relationship is approximately linear with a slope that depends on $\alpha_0$. The $R^2$ values quantify how well each linear regression fits the data.

To elaborate, the relationship between the spatial offsets on the galvo and the resulting slope of the phase ramp is detailed below. The modulated OPL of the reference arm can be given by $\Delta z \approx 2s\,\Delta\alpha$ [16–19] where $s$ is the off-center distance from the pivot axis, $\Delta\alpha$ is the galvo rotation angle relative to its central position, and Δz is the corresponding one-way geometrical path-length change (**Fig. 1A**). Now consider the phase change between each pair of adjacent line scans, i.e., the slope of the phase ramp, is

$$\phi = k(2\delta z) = \frac{4\pi\,\delta z}{\lambda} \approx \frac{8\pi s\,\delta\alpha}{\lambda} \qquad (1)$$

where $\delta\alpha$ is galvo's angular step size between successive positions, $\delta z$ is the corresponding one-way OPL change per scanning step, $k = \frac{2\pi}{\lambda}$ is the central wavenumber, and the factor of 2 accounts for the round-trip propagation in the reference arm. **Eq. (1)** shows that the slope of the phase ramp is proportional to both the galvo's angular step size, δα and the beam offset, $s$. By adjusting the galvo step size and beam offsets, we can select a phase-modulation slope for each reference arm so that the corresponding signals occupy distinct, nonoverlapping frequency bands in the transverse Fourier domain.

The above analysis and previous studies employing scan-induced phase modulation [16–19] all assume a central angle of incidence (AOI) at the galvo of 45° and an approximately linear relationship between galvo rotation and the induced phase modulation. In practice, achieving a 45° central AOI for every beam is difficult, particularly when multiple incident beams inevitably have small angular misalignments. Therefore, it is crucial to evaluate the validity of these assumptions under realistic experimental conditions. We did so by numerically simulating the exact geometric relationship between galvo rotation and the resulting OPL change:

$$\Delta z = s[\cot(\alpha_0 + \Delta\alpha) - \cot(\alpha_0)] \quad (2)$$

For a beam offset of s=1.5 mm, **Eq. (2)** gives the relationship between phase shift and galvo scan angle shown in **Fig. 1C**. We found the phase modulation slope strongly dependent on the central AOI, $\alpha_0$. For example, reducing $\alpha_0$ from 45° to 40° increases the slope magnitude by approximately 21%, whereas increasing it to 50° decreases the slope magnitude by 15%. Despite the change in slope, the phase shift remains approximately linear across a ±1.5° range of galvo positions, over the tested range of $\alpha_0$: $R^2 > 0.999$ for all tested $\alpha_0$ values (**Fig. 1C**). Smaller values of $\alpha_0$ produce steeper phase response. This is reassuring: practical alignment errors primarily change the slope of the phase modulation, with little phase nonlinearity.

It is also worth noting that the phase shift imposed by the reference-arm galvo (**Fig. 1C**) results from a scan-dependent change in reference-arm OPL, causing an apparent axial shift that varies across the transverse (slow-scan) direction of the B-scan. Because the galvo is scanned symmetrically about the central angle, this scan-induced OPL change is zero at the central A-scan, for all three reference arms. Thus, this effect should not affect axial-length measurements made at the center of the field of view; accordingly, no correction for this effect is applied when determining axial length.

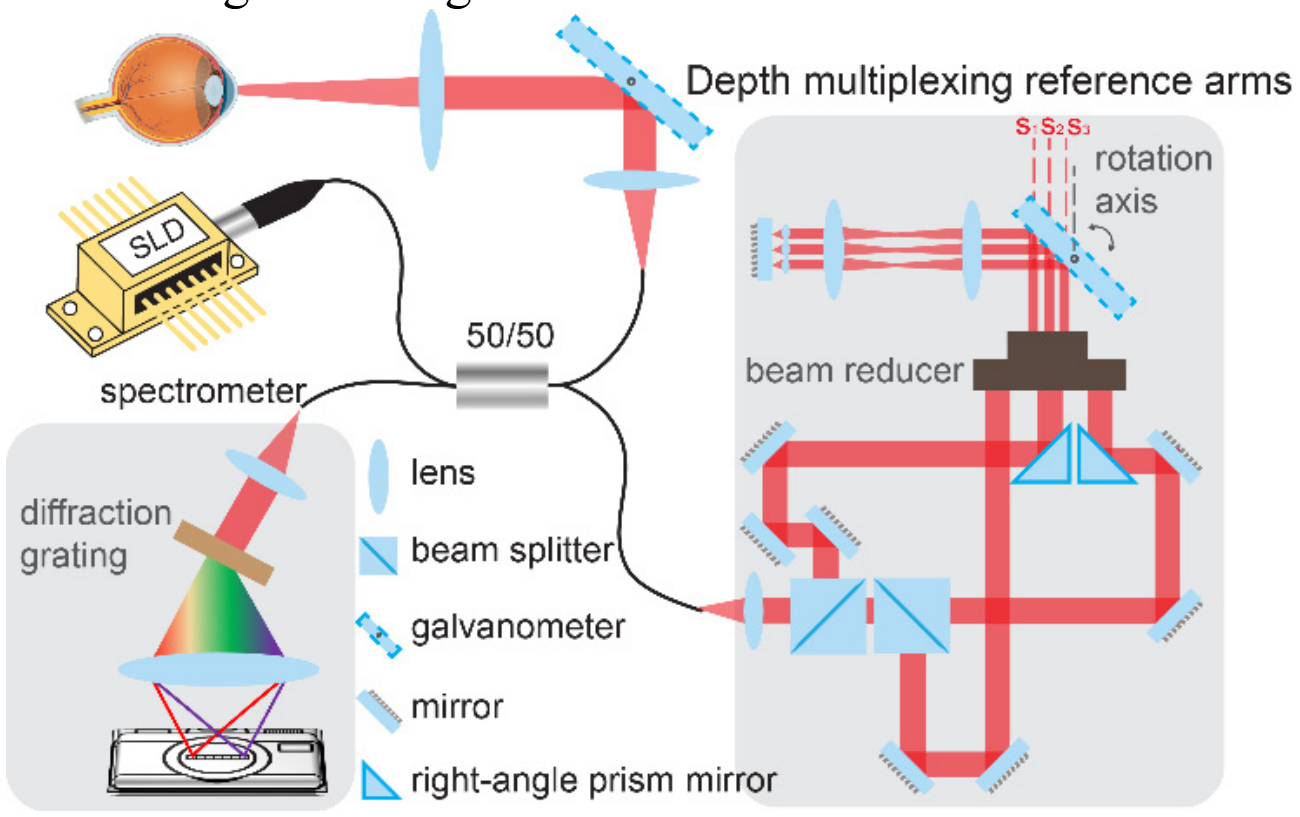

**Figure 2**. **Design of the depth-multiplexing spectral-domain optical coherence tomography (DM-SD-OCT) setup**. Three reference beams with different OPLs are directed onto the reference arm galvanometer at different lateral offsets, producing distinct phase modulation frequencies upon scanning. All three interference signals are collected simultaneously by a single spectrometer.

## B. Design and Implementation of the Optical Setup

In practice, each reference beam must be focused and collected by a separate lenslet (**Fig. 1B, Fig. 2**). This is because the reflected beam at the central AOI $\alpha_0$ should co-align with the optical axis of the corresponding lenslet, allowing the reflected beam to retrace its original path. A single lens cannot satisfy this condition for beams at different lateral positions. The lenslet array used in our system has pitches of 1.0 mm and 1.4 mm along its two orthogonal axes. We use the axis with a 1.0 mm pitch, requiring an integer number times of 1.0 mm center-to-center separation between adjacent beams.

The collimated beam diameter is approximately 3.25 mm ($1/e^2$ waist diameter), which is too large to accommodate these closely spaced beam positions without substantial overlap. A 5x beam reducer is therefore used to reduce the beam diameter before the beams reach the galvo scanner.

The entire system is built from commercially available off-the-shelf parts. The reference arm set consists of two cube beam splitters (BS005 and BS053, Thorlabs Inc.), a 5x beam reducer (GBE05-B, Thorlabs Inc.), a galvo scanner (GVS011, Thorlabs Inc.), two conjugation lenses (Thorlabs Inc.), a lenslet array (MLA1M2, Thorlabs Inc.) and a mirror (**Fig. 2**). The beam splitters divide the reference beam into three paths with distinct OPLs. The spectrometer is built using a low-cost line camera (Spyder3 1k, Teledyne DALSA) along with a holographic grating (HD 1800, Wasatch Photonics) and off-the-shelf lenses (Thorlabs Inc.).

## C. Demixing Pipeline

As each reference arm is assigned a distinct linear phase ramp along the transverse scan direction, signals from different reference arms occupy separate transverse spatial-frequency bands and can be demixed. As illustrated in **Fig. 3**, an image of a model eye (Rowe Technologies Inc.) obtained with a Zeiss IOL Master 700 (Carl Zeiss Meditec AG) is fed into a simulation pipeline to demonstrate the demixing procedure.

For demonstration purposes only, when each reference arm occupies $1/3$ of the full $\pi$ phase range in the quadrant, the phase shift for the three reference arms will be $\pi/6$, $\pi/2$, $5\pi/6$. A one-dimensional Fourier transform is first applied along the transverse axis of the simulated B-scan generated from the ground truth image (**Fig. 3B**). The frequency band associated with each reference arm is then isolated using a cropping window placed around each band, followed by an inverse Fourier transform along the transverse axis to recover the corresponding demixed spectral interferograms. Finally, an inverse Fourier transform along the spectral axis reconstructs the depth-resolved image for each channel (**Fig. 3C**), allowing the cornea, pupil, and retina signals to be independently recovered from a single multiplexed B-scan.

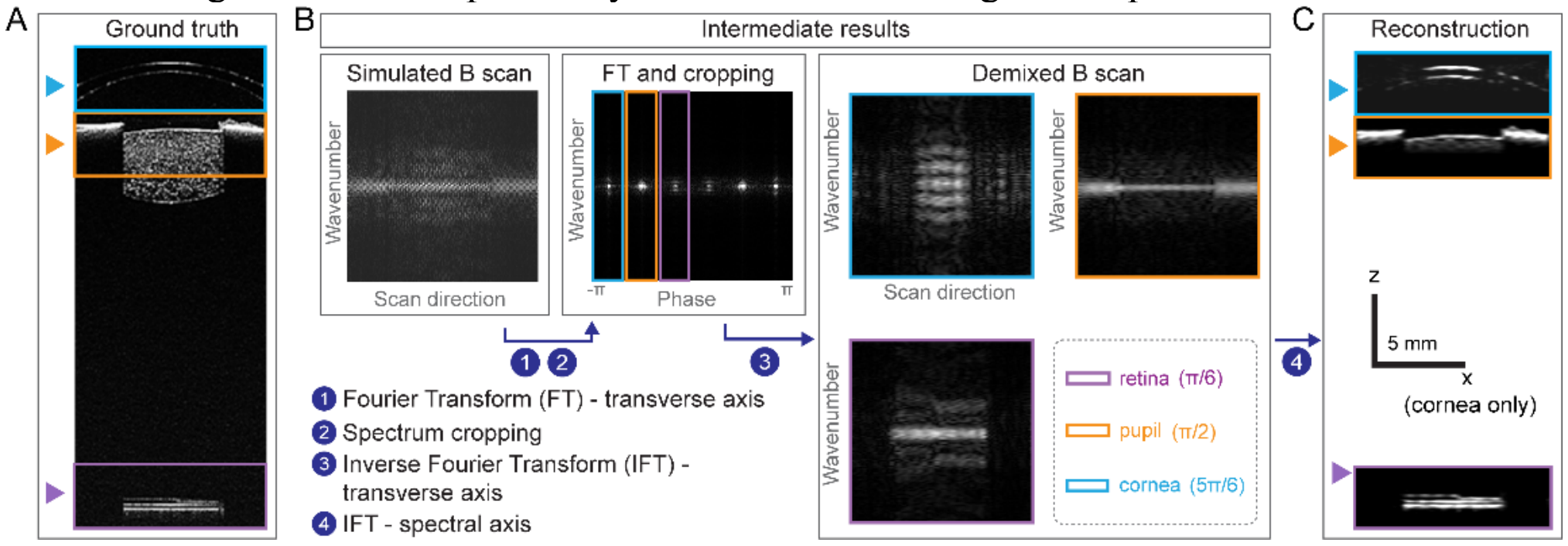

**Figure 3. Simulation and reconstruction pipeline for depth-multiplexed SD-OCT. (A)** Ground truth model eye image (obtained with a Zeiss IOL Master 700) used for simulation. Arrowheads: three reference arms target different depths within the eye. **(B)** Intermediate results of the demixing pipeline. A simulated frequency-encoded B-scan (cross-correlation term only) is first Fourier transformed along the transverse scan axis (1), separating the three depth channels into distinct transverse spatial-frequency bands. Each frequency band is then isolated by cropping (2), followed by an inverse Fourier transform along the transverse axis (3) to recover the demultiplexed spectral interferogram for each depth channel. **(C)** Inverse Fourier transform along the wavenumber axis (4) reconstructs the corresponding depth-resolved images.

## 3. RESULTS

### A. System Characterization

Signal fall-off and reference arm depth characterization were performed using a protected silver mirror as the sample. The home-built spectrometer provides a 0.11 nm spectral resolution, yielding a 1.4 mm imaging range with a 6 dB signal fall-off (**Fig. 4**). By moving the protected silver mirror axially with a 500 µm step size across a 40 mm range and fitting the measured image intensity to three distinct Gaussian curves, the depths of the three reference arms were found to be centered at 0 mm, 5.07 mm, and 33.56 mm in air, respectively (**Fig. 4**). With a 50:50 fiber coupler, 0.6 mW power into the eye, and a 50 µs camera exposure time (corresponding to a 20 kHz A-scan rate), we achieved a peak signal-to-noise ratio (SNR) above 90 dB at all three depths (**Fig. 4**). Alternatively, with a 90:10 fiber coupler (90% power goes to the reference arm) and 0.6 mW power into the eye, we achieved a peak SNR of 96 dB at a 27 µs exposure time (36 kHz A-scan rate, the maximum line rate allowed by the camera). The 50:50 coupler simplifies operation since its ports connect to either arm interchangeably and keep the beams in both arms visible for alignment. It achieved sufficient SNR for image acquisition and was used in all subsequent experiments. The axial resolution of our system is 14.4 µm. With a $1/e^2$ beam diameter of 3.25 mm, a focal length of 200 mm (focused in the middle of the crystalline lens), and an effective numerical aperture of 0.01, the lateral resolution of our system is 51 µm.

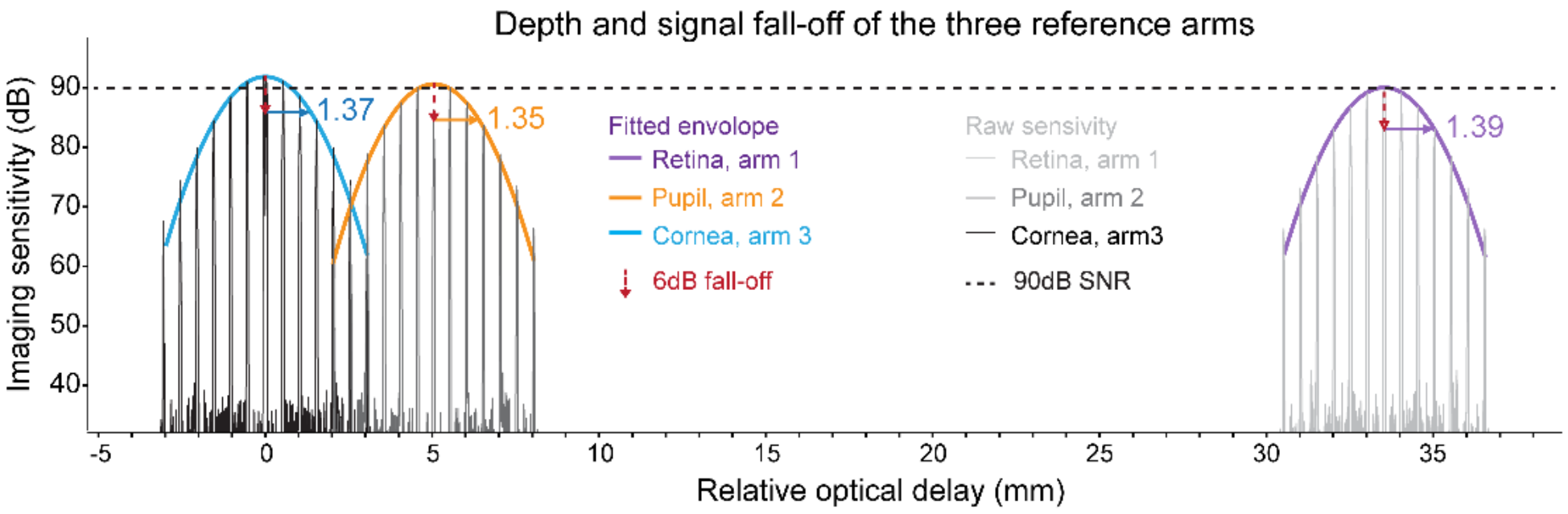

**Figure 4.** Characterization of the DM-SD-OCT system. Measured sensitivity roll-off and relative axial delay positions for the three reference arms. The raw sensitivity peaks are fitted to Gaussian envelopes tracking reference arm 3 (Cornea, blue) centered at 0 mm, arm 2 (Pupil, orange) centered at 5.07 mm, and arm 1 (Retina, purple) centered at 33.56 mm in air. The horizontal dashed line indicates that a peak SNR maintained above 90 dB across all three depths.

Although equal division of phase space among the three arms works well in simulation (**Fig. 3**), a practical concern prevents us from using a large phase slope. Because the exact initial angle of incidence ($\alpha_0$) and beam offset ($s$) are hard to measure accurately, we always determine the experimental phase shifts from the Fourier transform of the interferogram, following the procedure annotated in **Fig. 3B**. As expected from **Eq. (1)**, the phase-ramp slope is proportional to the galvo's angular step size (**Fig. 5A**). However, sensitivity measurements showed a significant fall-off at large phase slopes (**Fig. 5B**). The sensitivity fall-off can be interpreted as a consequence of fringe washout during each camera exposure. During image acquisition, the reference galvo was driven by a continuously varying voltage ramp synchronized with the camera line rate, rather than operated in a step-and-settle mode. As a result, the mirror continued to move throughout each exposure, continuously changing the reference-arm OPL and, consequently, the interferometric phase. Because the camera integrates the detected intensity over the finite exposure time, interference signals acquired at different phases are temporally averaged within each A-scan, washing out the interference fringes. This averaging reduces the effective fringe contrast and therefore the reconstructed OCT signal amplitude. The effect becomes stronger at larger phase-ramp slopes, which cause greater phase changes and fringe motion during each exposure, leading to more pronounced washout and a larger reduction in sensitivity. To minimize overlap among the signals while limiting the sensitivity fall-off at large phase slopes, we experimentally optimized the phase modulation settings and adopted phase slopes of π/5, 2π/5, -2π/5 for the cornea, pupil, and retina arms, respectively. The galvo's angular step size was approximately 35 µrad per scan step.

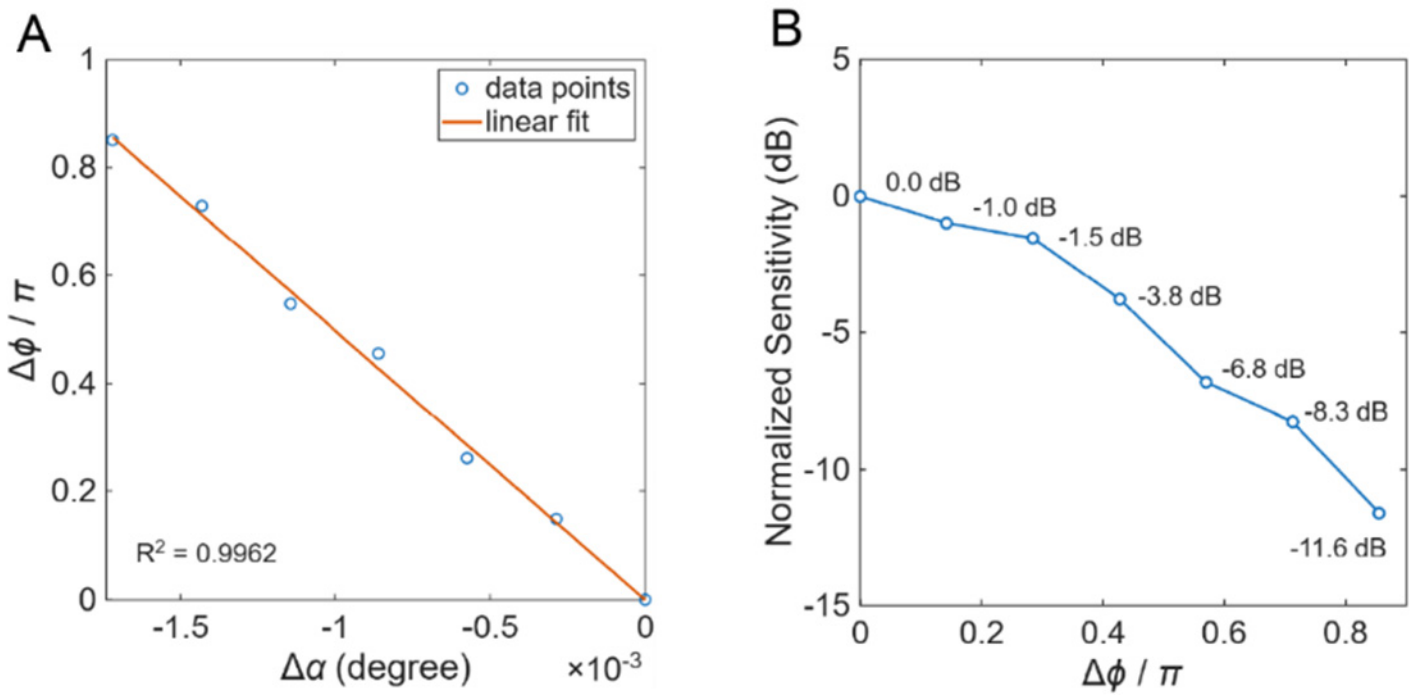

**Figure 5. Experimental determination of phase modulation slope and associated signal fall-off.** (A) Observed linear relationship between the phase slope and the galvo's angular step size. (B) Measured normalized sensitivity as a function of the slope of the phase modulation

## B. Validation of Depth Multiplexing and Demixing Using a Model Eye

To evaluate the performance of the DM-SD-OCT system, initial testing was performed using a model eye (Rowe Technologies Inc.). A custom Graphical User Interface (GUI) developed in C++ provided real-time visual guidance by displaying reconstructed images at the frame acquisition rate. The demixing algorithm to resolve the multiplexed depths was also integrated into the real-time processing pipeline. This high-speed pipeline allowed for the instantaneous real-time adjustment of the eye position during experiments.

As shown in **Fig. 6**, our system successfully captured information from the cornea, pupil, and retina of the model eye. After removing the DC term and the autocorrelation term (the interference from the cornea and pupil arms) by subtracting the B-scan without the sample in place, we obtained a clean B-scan containing only the sample-reference interference, i.e., the cross-correlation terms. If a single inverse Fourier transform was applied to the B-scan, all the structures can be reconstructed but with significant overlap (**Fig. 6A**). In contrast, with our demixing pipeline applied, we are able to separate signals from all three distinct depths without noticeable overlap (**Fig. 6B**). A single frame was sufficient to reveal all structures, although a 10-frame average improved the signal to noise ratio (SNR) (**Fig. 6C**). A fitting was performed to identify the dimensions of the model eye: curved boundaries were determined using a second-order polynomial fitted using three manually selected points along each boundary. For the retinal image, the fitting was defined by a straight line aligned with the first prominent high-intensity reflection from the retinal surface. Our results match the measurements using Zeiss IOLMaster 700 (**Table 1**).

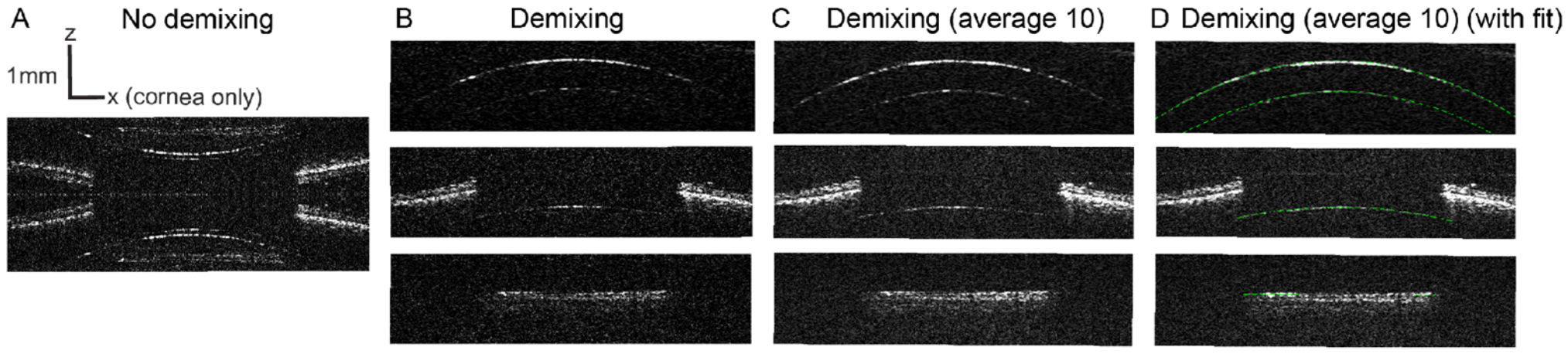

**Figure 6.** Experimental validation in a model eye, with π/5, 2π/5, -2π/5 phase slopes. **(A)** DM-SD-OCT result without frequency demixing, using 3 reference arm depths targeting the cornea, pupil and retina, respectively. **(B)** Frequency demixed images from a single frame acquisition. **(C)** Frequency demixed images averaged from 10 frames. **(D)** Averaged frequency demixed images with fitting.

| | DM SD-OCT | Zeiss IOL Master |
|---|---|---|
| Axial length (AL) | 22.83mm | 22.85mm |
| Central corneal thickness (CCT) | 0.51mm | 0.50mm |
| Aqueous depth (AQD) | 3.15mm | 3.16mm |
| Anterior chamber depth (ACD) | 3.66mm | 3.66mm |

**Table 1.** Model eye measurement comparison between DM-SD-OCT and Zeiss IOL Master 700.

### C. *In vivo* Ocular Biometry in Human Subjects

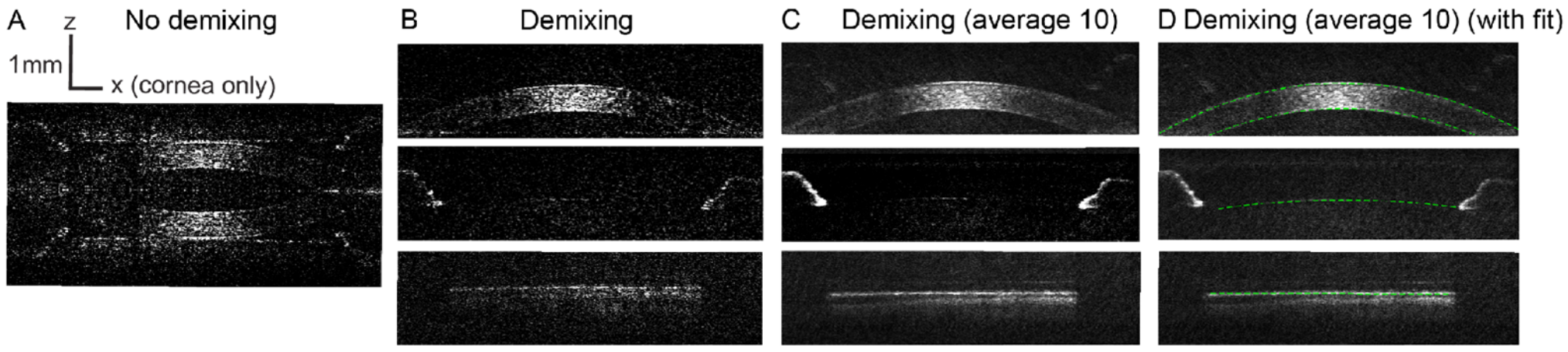

**Figure 7.** Experimental validation in a human eye *in vivo*. **(A)** DM-SD-OCT result without frequency demixing, using 3 reference arm depths targeting the cornea, pupil and retina, respectively. **(B)** Frequency demixed images from a single frame acquisition. **(C)** Frequency demixed images averaged from 10 frames. **(D)** Averaged frequency demixed images with fitting.

The system's clinical utility was further validated through *in vivo* imaging of three healthy human subjects: one emmetropic individual and two myopic individuals with different levels of myopia. The University of California Berkeley Institutional Review Board approved this research, and subjects signed an informed consent form before participation. Subjects were seated and initially positioned to approximate the alignment used for the model eye. Subsequent fine adjustment of the lateral shift and zero-delay position was then performed guided by the real-time GUI display. A representative B-scan captured during this acquisition process is shown in **Fig. 7A**. Following the demixing process, signals from different axial regions were separated, allowing the corneal, lens, and retinal layers to be resolved in depth (**Fig. 7B**). To extract essential biometric data, a curve shape-fitting algorithm was employed to determine four key parameters: axial length (AL), central corneal thickness (CCT), aqueous depth (AQD), and anterior chamber depth (ACD), which are important indices for monitoring eye growth.

| | Subject 1 | | Subject 2 | | Subject 3 | |
|---|---|---|---|---|---|---|
| | DM SD-OCT | Zeiss IOL Master | DM SD-OCT | Zeiss IOL Master | DM SD-OCT | Zeiss IOL Master |
| Axial length (AL) | 26.75mm | 26.79mm | 24.34mm | 24.36mm | 22.65mm | 22.67mm |
| Central corneal thickness (CCT) | 0.53mm | 0.54mm | 0.55mm | 0.55mm | 0.56mm | 0.54mm |
| Aqueous depth (AQD) | 3.12mm | 3.13mm | 2.64mm | 2.63mm | 2.83mm | 2.85mm |
| Anterior chamber depth (ACD) | 3.65mm | 3.67mm | 3.19mm | 3.18mm | 3.39mm | 3.39mm |

**Table 2.** Human eye measurement comparison between DM-SD-OCT and Zeiss IOL Master 700.

The OPLs for each ocular feature were first extracted from the demixed images relative to their respective reference arms. To determine the absolute OPL for each ocular segment, we integrated these values with the calibrated axial separations between the three reference arms. These absolute OPL values were then converted into physical lengths after accounting for the refractive properties of the ocular media [20].

As shown in **Table 2**, our results demonstrated strong agreement with measurements obtained from the commercially available Zeiss IOLMaster, suggesting our DM-SD-OCT system provides a robust, cost-effective platform for comprehensive ocular biometry without compromising diagnostic accuracy.

## 4. DISCUSSION

In summary, we proposed a new depth-multiplexing method for SD-OCT using a single phase modulation unit that requires no additional detectors. By encoding each reference arm with a distinct phase ramp and demixing the depths computationally in real time, the system recovers the cornea, pupil, and retina from a single acquisition, and its biometric measurements agree closely with a commercial swept-source biometer. Because the entire setup is built from off-the-shelf components and tolerates a low-cost, low-speed line camera, it offers an affordable alternative for full-eye-length measurement, and the approach applies to any SD-OCT devices where an extended imaging depth is desired.

The principal limitation is the reduced dynamic range of the spectrometer when multiple reference arms are captured simultaneously on the same sensor although this was not a practical concern in our experiments. A second constraint is that demixing relies on phase stability across the transverse scan; mechanical vibration broadens the frequency bands and degrades separation, which sets a practical bound on how many depths can be multiplexed.

Taken together, these results demonstrate that full-eye-length biometry can be achieved on a low-cost SD-OCT platform without the swept source and high-bandwidth acquisition that make conventional biometers expensive. Such a system could broaden access to axial-length monitoring in settings where cost is a barrier, including the pediatric myopia screening that motivates much of the current demand for ocular biometry.

**Acknowledgement.** We thank Dr. John G. Flanagan for inspiring discussions throughout this project. This work was funded by the Center for Innovation in Vision and Optics (CIVO) at the University of California, Berkeley.

**Disclosures.** The authors declare no conflicts of interest.

**Data and code availability.** Data and code will be available upon reasonable request to the corresponding author.

**References.**